\documentclass[prl,twocolumn,showpacs,preprintnumbers,amsmath,amssymb,floatfix]{revtex4}
\usepackage{epsfig}
\usepackage{graphicx}
\usepackage{amssymb}
\usepackage{color}

\begin{document}

\title{On the complexity of controlling quantum many-body dynamics}

\author{T.~Caneva$^{1,2}$}
\author{A.~Silva$^3$$^,$$^4$}
\author{R.~Fazio$^5$}
\author{S.~Lloyd$^6$}
\author{T.~Calarco$^1$}
\author{S.~Montangero$^1$}
\affiliation{$^1$Institut f\"ur Quanteninformationsverarbeitung, 
Universit\"at Ulm, 89069 Ulm, Germany.}
\affiliation{$^2$ ICFO-Institut de Ciencies Fotoniques, Mediterranean Technology Park, 08860 Castelldefels (Barcelona), Spain.}
\affiliation{$^3$ SISSA -- International School for Advanced Studies, via Bonomea 265, 34136 Trieste, Italy.}
\affiliation{$^4$ Abdus Salam ICTP, Strada Costiera 11, 34100 Trieste, Italy.}
\affiliation{$^5$ NEST Scuola Normale Superiore \& Istituto di Nanoscienze CNR, Piazza dei Cavalieri 7, I-56126 Pisa, Italy.}
\affiliation{$^6$ Department of Mechanical Engineering, Massachusetts Institute of Technology, Cambridge, MA 02139, USA.}

\begin{abstract}
We demonstrate that arbitrary time evolutions of many-body quantum systems can be reversed  even in cases when only part of the Hamiltonian can be controlled. 
The reversed dynamics obtained via optimal control --contrary to standard time-reversal procedures-- is extremely robust 
to external sources of noise.  {We provide a lower bound on the control complexity of a many-body quantum dynamics in terms of 
the dimension of the manifold supporting it, elucidating the role played by integrability in this context.}
\end{abstract}

\maketitle

In recent years,  fast progress on the understanding of non-equilibrium dynamics of many-body quantum systems has 
been spurred by unprecedented opportunities offered by  cold atom {quantum simulation experiments~\cite{Bloch2008}. 
At the same time, powerful numerical tools~\cite{Schollwock2011} have made it possible  to investigate the}
out-of-equilibrium dynamics of many-body quantum systems and to compare theoretical results 
with experimental data obtained in {highly controlled and tunable systems.} Many interesting 
situations have been already experimentally investigated so far~\cite{Polkovnikov_RMP11} including 
(just to give a few examples) quench dynamics~\cite{Cheneau2012}, 
thermalization~\cite{Trotzky2012}, quantum phase transition dynamics~\cite{Simon2011},  {and the effect 
of periodic}  perturbations~\cite{Chen2011, Lignier2009}  both in fermionic and bosonic systems~\cite{Bloch2012}.  

{Given the ability to engineer a large class of  Hamiltonians, the challenge for the future 
will be to be able to engineer the full time evolution of the many-body quantum state by shaping the time-dependence of few 
control parameters, e.g.  coupling constants and external fields.
This ability, beyond the bounds of possibility until a few years ago, paves the way for the 
realisation of many-body state engineering, with optimal control techniques~\cite{walmsley2003} 
emerging as the ideal tool to use.  }

\begin{figure}
\epsfig{file=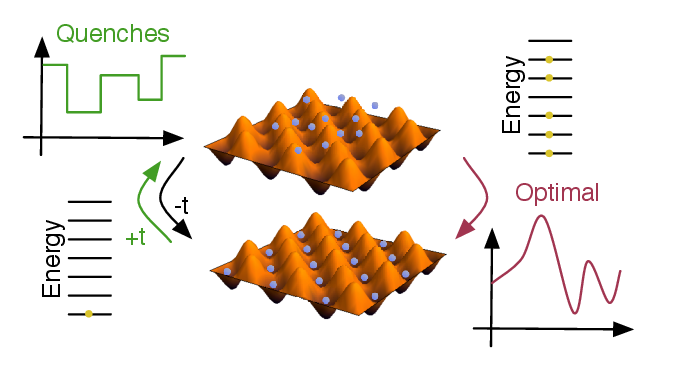,width=8cm,angle=0,clip=}
\caption{(Color online) Dynamical scheme to optimally reverse the system dynamics: a system in the ground state is taken out of equilibrium by multiple random quenches. 
Reversing the dynamics can be obtained in general via a time-inversion or by solving an optimal control problem.}
\label{scheme}
\end{figure}

{Quantum optimal control, routinely used in many areas of science~\cite{walmsley2003},  has been applied only recently to 
quantum many-body systems, e.g.
for the state preparation of strongly interacting cold atoms in optical lattices and spin systems~\cite{CRAB},  
to analyse the crossing of a quantum critical point~\cite{caneva2011} or to the cooling 
of Luttinger liquids~\cite{rahmani}.}
The theoretical study and experimental implementation of optimal control strategies to quantum 
many-body systems poses in turn a number of important questions. 
While it has been shown how quantum optimal control can drive a few-body system up to its quantum 
speed limit~\cite{caneva2009,bason2012},  
it is important to understand to which extent is it possible to control a 
quantum many-body system. Which resources are needed in terms of complexity,  
in particular in connection to the integrable/chaotic nature of the system under investigation? 
And how efficient and robust will the resulting control strategy be?
 
Answering these questions would bring together in a new perspective thermodynamics, optimal control,  
and complexity theory  {thus} paving the way to further developments and investigations.
In particular, an interesting related issue is the reversibility of closed many-body 
quantum systems dynamics, which
might have intriguing consequences on a fundamental problem in physics, i.e. the emergence of 
the arrow of time. Indeed, one can revert the dynamics of a quantum system 
by inverting the time propagator, as it is typically done 
in spin-echo experiments~\cite{tannoudji2011}. This procedure is however a 
highly non trivial task in a general many-body quantum system and requires an enormous accuracy in the 
knowledge of the history of the dynamical process and of the control field: 
the smallest deviation from the exact path inversion has dramatic 
consequences~\cite{Polkovnikov_AP11}. Moreover, very few systems are 
amenable to such operations, since quantum systems are typically only partially
tunable. In practice, reversing a complex time-evolution of a many-body quantum system is 
commonly believed to be unfeasible.

Aim of this article is to study the limits to optimal control dictated by the complexity (to be  
properly defined later) of a partially tunable quantum system. The result of our investigation is a qualitative and quantitative
characterisation of our ability to drive a many-body quantum system from a given input to a predetermined final state.
How do we accomplish this task? The idea is simple: Given an Hamiltonian depending on certain couplings, an initial
and a final state (reachable by the system during its evolution), we would like to see what are the resources needed 
by  optimal control to dynamically  connect the two states. This approach is meaningful if the input and target states 
represent a worst-scenario case. We define it by considering two states having maximum difference in the diagonal 
entropy. The latter has been introduced in~\cite{feldman03,Polkovnikov_AP11}  to characterize non-equilibrium quantum 
evolution. Here it will be used to probe the limits of optimal control.
In practice, starting from the ground state, we dynamically generate a random state with maximum diagonal entropy
and then we look for a strategy to drive back the system into the initial state (see Fig.~\ref{scheme}). 
{Incidentally, we notice that this protocol also provides a general scheme to drive the system between 
any two (dynamically connected) states. Indeed given any initial and target states it is possible to drive the system between them   
concatenating two optimal transformations:  initial to ground state and subsequently ground to the target one.  
In conclusion, if we quantify the complexity of the proposed protocol we are able to quantify the complexity of 
any state-to-state optimal control transformation.

Once having shown that a  reversing  protocol based on  optimal control exists and it is
robust against errors, we then address the problem on a more general ground, 
elucidating the relationship between the complexity of the control field (band-width)
and the dimension $D_m$ of the manifold in which the dynamics occurs, 
{providing an informational lower bound for it.} 
These results show that the intrinsic complexity of {controlling many-body dynamics and in particular of }
inverting time evolutions rests on the tensor-product structure of many-body quantum system Hamiltonians, that is, 
on the fact that in a non-integrable system the 
band-width necessary to invert the system dynamics (i.e. the arrow of time) scales exponentially with the system size.

\begin{figure}
\epsfig{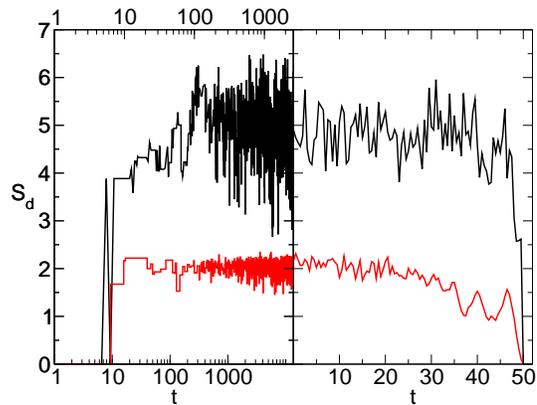}
\caption{(Color online) Diagonal entropy $S_d$ for the Ising model (black upper line) and 
for the LMG model (red lower line) as a function of time during the
disordering procedure (left side of the figure, logarithmic scale
for the time) and during the optimization (right side of the figure, 
linear scale for the time) with
$N=20$, $T_{max}=100/\Delta $, where $\Delta$ is the critical gap. Time is in units of J.}
\label{models_sd_vs_t:fig}
\end{figure}

\emph{Optimal reversed dynamics}-- 
{The program outlined above can be implemented by} the following dynamical scheme (see Fig.~\ref{scheme}):
\emph{i}) we first initialize the system in its ground state; 
\emph{ii}) we then apply a disordering quench process, generating an highly excited state; 
\emph{iii}) we finally steer the system back into 
the initial state 
either using an optimized protocol or time-reversal.

{For our numerical computations we consider  systems described by the 
spin-1/2 Hamiltonian:}
\begin{eqnarray}
H=-\sum _{i,j}J_{ij}\sigma_i ^x \sigma_{j} ^x 
    -\Gamma (t)\sum _{i}^N \sigma_i ^z -J_x\sum _{i}^N \sigma_i ^x,
\label{Ising_ham_xz:eq}
\end{eqnarray}
For vanishing longitudinal field ($J_x=0$), this Hamiltonian has two obvious integrable limits,
the quantum Ising chain in transverse field, where $J_{ij}=J\delta _{i,i+1}$, 
and the infinite-range quantum Ising model (or Lipkin-Meshkov-Glick (LMG) model~\cite{Lipkin_NP65}) 
when $J_{ij}=J/N$ for $i<j$ ($J_{ij}=0$ otherwise). In the presence of a longitudinal field 
($J_x \ne 0$), the quantum Ising
chain loses its integrability~\cite{Karthik}, apart from at the critical  point (in the scaling 
limit)~\cite{Zamo}. From now on we set $\hbar=1$ and time is expressed in units of $J$.

{The dynamics of the LMG model takes place}
in the subspace generated by the Dicke states $|\mathcal{S},\mathcal{S}_z\rangle$,
where $\mathcal{S}$ is the conserved total angular momentum
and $-\mathcal{S}\leq\mathcal{S}_z\leq \mathcal{S}$ are the allowed possible projections along the 
$z$-axis~\cite{Botet_PRB83}. 
The ground state of the Hamiltonian belongs to the subspace with $\mathcal{S}=N/2$
and in the following we are working within this dynamically accessible subspace, composed by 
$D_{LMG} \sim N/2$ Dicke states (corresponding to the allowed values of $\mathcal{S}_z$ with the correct parity).
The Ising chain in transverse field for $J_x=0$ can in turn be solved exactly through
the Jordan-Wigner transformation mapping the spins onto free fermions~\cite{Lieb_AP61}, 
the dimension of the Hilbert space is $D_{I}=2^{N/2}$ due 
to parity conservation for the number of fermions obtained with the 
Jordan-Wigner transformation.\\
For both models, we prepare the system in the ground state $|\psi (0)\rangle= |GS\rangle$
at large value of the driving field $\Gamma$ -- a fully polarized spin state along the positive 
$z$-axis. 
We then drive the system out of equilibrium performing a repeated quench between two values
$\Gamma _1$ and $\Gamma _2$ of the control field $\Gamma$; 
each quench lasts a random waiting time $T_{max}\cdot r_i$, where $T_{max}$ 
is the maximum allowed waiting time and $r_i\in [0,1]$ is a uniformly distributed random 
number~\cite{Polkovnikov_AP11}. 
We quantify the complexity of the out-of-equilibrium state via the diagonal entropy $S_d= \sum p_i \log p_i$, 
the entropy of the populations $p_i$ of the density matrix of the system  in the instantaneous 
Hamiltonian eigenbasis~\cite{feldman03,Polkovnikov_AP11}. In the LMG model,  
the maximal achievable diagonal entropy scales \emph{logarithmically} with the size, 
$S_d^{LMG}\sim\log (N/2+1)$, while in the Ising chain it scales 
\emph{linearly} with the size $S_d^{I}\sim (N/2)\log(2)$.  
Indeed the dimension of the accessible Hilbert space is $D_{LMG} \sim N/2$  
and $D_{I}=2^{N/2}$ respectively.
We verified that after a sufficiently large number of cycles the average $S_d$ produced with 
this disordering procedure is approximately independent of the
amplitude $|\Gamma_1 -\Gamma _2|$ and of the waiting time between two consecutive quenches.\\
\begin{figure}
\epsfig{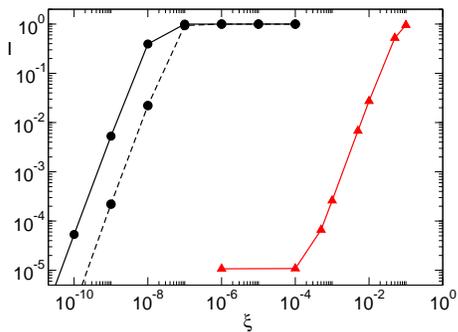}
\caption{(Color online) Infidelity $I$ as a function of the intensity of the noise $\xi$ in the 
Ising model for $N=20$, with 
$\Gamma(t)=\Gamma (1+\xi \cdot r(t))$,
where $\Gamma$ is the driving field in the absence of noise and $r(t)$ is a 
random variable uniformly distributed in $[-1,1]$. Black circles continuous (dashed) line: 
exact reverse dynamics $T\sim 2\cdot 10^3$ ($T\sim 90$); red triangles continuous line: 
optimized reverse $T\sim 50$. Time is in units of J.}
\label{trev_opt_inf_vs_noise_ising:fig}
\end{figure}
Finally, we use optimal control to drive the system from the out of equilibrium state back to 
the initial state $|\psi (0)\rangle$, in a given time $T$ to obtain the final state  $|\psi (T)\rangle$. 
The optimization is implemented through the Chopped Random Basis (CRAB) technique~\cite{CRAB}: 
after making a guess for a possible return path $\Gamma_0(t)$, we introduce a correction
of the form $\Gamma(t) = \Gamma_0(t) f(t)$, where the function 
$f(t)$ is expressed as a truncated Fourier series, i.e.
$f(t) = 1+ \sum_{k}  A_k sin(\nu_kt) + B_k cos(\nu_kt)/\lambda(t).
$
Here, $k = 1, . . . ,n_f$, $\nu_k = 2\pi k(1 + r_k)/T$ are �random-
ized� Fourier harmonics, $T$ is the total time evolution,
$r_k \in [0,1]$ are random numbers with a flat distribution,
and $\lambda(t)$ is a normalization function to keep the initial
and final control pulse values fixed. The optimization
problem is then reformulated as the extremization of a
multivariable function $F({A_k}, {B_k}, {\nu_k)}$, which can
be numerically performed with a suitable method~\cite{CRAB}.

In Fig.~\ref{models_sd_vs_t:fig}
typical results of this procedure are shown for the LMG model (red line) and the Ising model 
(black line): in the left half of the picture the disordering process is applied and 
the diagonal entropy increases reaching an average maximum value ($S_d^{LMG}\sim\log (N/2)$ 
 and $S_d^I\sim N/2\log (2)$). 
In the right half of the picture the optimization phase is shown:  
even though the  Hamiltonian is only partially tunable
(i.e. we do not allow for a sign reversal of all couplings) the control is able to steer the system 
towards the desired initial stationary state, reducing $S_d$ to zero, without the need for 
information on the history of the disordering process.
We verified that this holds for different system sizes and different
total times  $T$ (data not shown).

As discussed in Ref.~[\onlinecite{Polkovnikov_AP11}], driving a system back to its initial state 
by a full time-reversal of a protocol is a procedure extremely sensitive to very small noise
perturbations.
It is therefore natural to compare the effects of noise on a naively time-reversed protocol
to those on an optimised CRAB protocol.  
In order to do so, we will consider a perturbed protocol $\tilde \Gamma(t)= \Gamma(t) [1+ r(t) \xi]$, where 
$\Gamma(t)$ is the original protocol (including a sign change of the couplings for 
the time reversed one), $r(t)$ is a random variable uniformly distributed in 
$[-1,1]$, and $\xi$ is the intensity of the noise. 
In Fig.~\ref{trev_opt_inf_vs_noise_ising:fig}
we plot the final infidelity $I$ as a function of $\xi$
for the exact reverse dynamics (black circles continuous line)
and for the optimized reverse dynamics (red triangle continuous line). 
The robustness of the optimized protocol emerges strikingly: a noise
more than six orders of magnitude more intense is
needed to affect the optimal protocol to an extent comparable to the time-reversed one. 
In order to verify that the stability of the optimized protocol is not simply due to 
the reduced total evolution time, we repeated the analysis with a time-reversed protocol lasting 
a time comparable to that of the optimized process: the minimal time to obtain a maximal 
entropy state is around $T\sim 90$ (black circles dashed line), to be compared with 
the duration of the optimal process, $T\sim 50$. The results shown in 
Fig.~\ref{trev_opt_inf_vs_noise_ising:fig} confirm
that the optimal protocol is intrinsically more stable than the time-reversed one, making 
it an excellent candidate for experimental implementations.

\begin{figure}
\epsfig{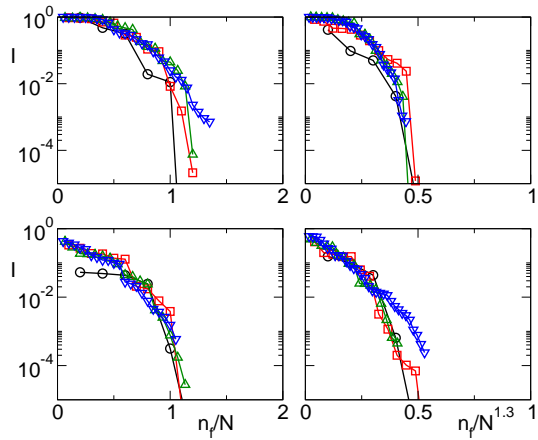}
\caption{(Color online) Infidelity $I$ as a function of the rescaled number of frequencies 
$n_f/N^{\alpha}$ in the Ising (upper panels) and LMG (lower panels) models, 
for the two transitions $|C\rangle\rightarrow|GS\rangle$ (left panels)
and $|M_S\rangle\rightarrow|GS\rangle$ (right panels) 
with $T=50$ (ising) and $T=100$ (LMG) and $N=10,20,30,40$ (black circles, red squares, 
blue triangles, green triangles respectively). Time is in units of J.}
\label{inf_vs_rescal_freq:fig}
\end{figure}
\emph{Control complexity}-- 
{We would like now to characterize the complexity of such }
a generic state-to-state conversion problem. Let us first give an operational definition of 
complexity:
within the CRAB algorithm we will measure the complexity in terms of the number $n_f$ of 
Fourier components needed to solve the optimisation problem up to a certain target infidelity. 
Intuitively transformations from a state with maximal diagonal entropy (i.e. completely 
delocalized in 
phase space) to one with low diagonal entropy (i.e. well localized) are expected to be more 
difficult than
those between localized states or between adiabatically connected states. 
It turns out however that the complexity of an optimisation protocol depends only weakly on 
the choice of initial and final states. Let us illustrate this considering two 
different state-to-state transformations:
from a maximal entropy state to the ground state ($|M_S\rangle\rightarrow|GS\rangle$) and 
from an eigenstate at the center of the spectrum to the ground state 
($|C\rangle\rightarrow|GS\rangle$), for both the Ising model and the LMG model. 
%
In Fig.~\ref{inf_vs_rescal_freq:fig} we show the final infidelity 
for different system sizes $N$ as a function of the number of frequencies $n_f$, 
at fixed total time $T$.
In all cases considered, the infidelity decays exponentially 
with the rescaled number of frequencies $ n_f/B(N) =n_f/N^\alpha$, 
showing a very similar behavior for both states in both models:
indeed we have 
%
$I\sim g(n_f/N^{\alpha})$, 
%
$g(x)$  being a scaling function of the form $\exp (- x^\eta)$, with $5>\eta >1$
and $1 <\alpha < 1.5$.
%
%
The first interesting feature emerging from our analysis is that within each model
the two transformations $|M_S\rangle\rightarrow|GS\rangle$ and 
$|C\rangle\rightarrow|GS\rangle$ approximately present the same complexity: $\alpha$
is only slightly larger for the $|M_S\rangle\rightarrow|GS\rangle$ conversion.
This result can be explained by the fact that the states $|C\rangle$ and
$|GS\rangle$ are not trivially connected through the Hamiltonian, 
although they are both localized with respect to  $H[\Gamma]$
for $\Gamma\gg 1$. In practice also in this case the transformation is performed by first 
spreading the state onto the whole Hilbert space and then recombining  the different
amplitudes into the desired state.
Such an operation requires approximately the same complexity as the state-to-state 
conversion between the maximally spread state and the ground state, 
$|M_S\rangle\rightarrow|GS\rangle$.
The second feature is instead emerging from the comparison between the two different
models: the complexity scales approximately linearly with the size for both the LMG
and the Ising model. 

\begin{figure}
\epsfig{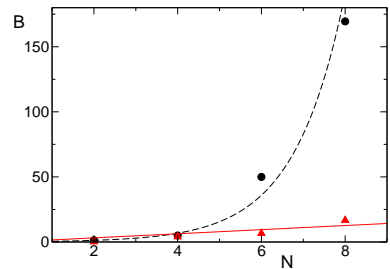}
\caption{(Color online) 
Dimensionless decay rate $B$ as a function of the size $N$ for the Ising model with $J_x\neq 0$ (black circles) 
and $J_x= 0$ (red triangles).}
\label{ising_long:fig}
\end{figure}
{\it Lower bound}-- The previous arguments suggest that 
the number of frequencies $n_f$ required in the optimal control pulse to achieve full control 
is not strongly dependent on the initial and final states but rather of 
the dimension of the manifold supporting the dynamics, namely  $D_m(N)$. 
{For the LMG model the linear scaling with size of manifold supporting the
dynamics, $D_m(N)\sim N$, is obtained by exploiting directly global symmetries
of the system; for the Ising model in absence of longitudinal field
$D_m(N)\sim N$ derives from the simplification introduced through
the Wigner-Jordan transformation~\cite{Lieb_AP61}.}
{Indeed we can provide a lower bound on the complexity of the optimisation task with 
the following information-theoretical argument:
the amount of information required to specify a state within an ball of radius $\epsilon$ in a 
$D_m$ dimensional Hilbert space is given by $b_\epsilon = \log_2 (1/ \epsilon^{D_m})$. 
A control field carries $b_f = T \cdot \Delta\Omega \cdot k_s = n_f  k_s$ bits of information 
where $T$ is the total length of the signal, and $\Delta\Omega$ and $k_s$ its bandwidth 
and bit depth respectively.  
The simplest control field that uniquely determines the goal state within the desired 
$\epsilon$-ball --and thus can drive the system from a reference state to the goal state-- 
has to carry at least the same information content, i.e. 
$b_\epsilon < b_f$. 
Solving for $\epsilon$, one founds the lower bound
\begin{equation}
\epsilon > 2^{-\frac{n_f k_s}{D_m}},
\label{bound}
\end{equation}
which implies that to keep a constant error the bandwidth of the control field 
should scale at least like $n_f \simeq D_m(N)$, as verified in the previous numerical 
optimizations (Fig.~\ref{inf_vs_rescal_freq:fig}).}
%
%
{
We finally consider the Ising model 
in the presence of a longitudinal field $J_x\neq 0$ in Eq.~\eqref{Ising_ham_xz:eq}.
In this case, we expect that the optimization complexity should increase drastically
since the dimension of the effective manifold supporting the dynamics $D_m(N)$
now scales exponentially with $N$.}
We performed simulations for both cases, $J_x=0$ (integrable system)
and $J_x\neq 0$ (non integrable system), analyzing the behaviour of the infidelity as a function
of the number of frequencies for systems of different sizes. 
In Fig.~\ref{ising_long:fig} we show the fitted decay rate values $B(N)$ as a function of the 
size $N$ for $J_x\neq 0$ (non-integrable, full circles) and $J_x= 0$ (integrable, empty triangles). 
Despite the fact that, due to the exponentially growing 
Hilbert space, we are now restricted to small sizes $2<N<8$, 
as clearly shown by the fits, the rate $B(N)$ for the integrable model ($J_x= 0$) 
scales linearly with the size, while for the non integrable model 
($J_x\neq 0$)  it scales as an exponential with $N$, that is $B(N) \propto D_m(N)$.
Interestingly,  in the case $J_x=0$, optimal control complexity scales approximately linearly 
even though those results have been obtained  by
simulating the Ising model in the full exponential Hilbert space without using the Wigner-Jordan 
transformation. In this sense, optimal control complexity appears not to be influenced by the simulation 
{details and thus CRAB control might be very effective in any integrable system.} 

{\emph{Conclusions}-- In this work we explored the limits of optimal control 
of the dynamics of a many-body quantum system.  We showed that  optimal control is able 
to reverse the dynamics of many-body quantum systems, 
effectively reducing the quantum entropy generated with strongly disordering 
processes and furthermore  that it might be possible to optimally reverse the system dynamics 
even in cases in which an exact reverse evolution cannot be realized.  
We demonstrated that the optimized reverse dynamics is extremely
robust against external sources of noise. Finally, we {provide a lower bound on the optimization complexity}
establishing a relationship between 
optimization complexity and integrability of the considered system. 
}

We acknowledge support from the EU through SIQS \& PICC, SOLID, from the German Research 
Foundation (SFB/TRR21), and the BW-grid for computational resources.
AS is grateful to KITP for hospitality. This research was supported in part by
the National Science Foundation under Grant No. NSF
PHY11-25915.



\bibliographystyle{apsrev}

\end{document}